\documentclass{article}
\usepackage[utf8]{inputenc}
\usepackage[T1]{fontenc}
\usepackage{xcolor}
\DeclareUnicodeCharacter{2080}{\textsubscript{0}}
\DeclareUnicodeCharacter{2081}{\textsubscript{1}}
\DeclareUnicodeCharacter{2082}{\textsubscript{2}}
\DeclareUnicodeCharacter{2083}{\textsubscript{3}}
\DeclareUnicodeCharacter{2084}{\textsubscript{4}}
\DeclareUnicodeCharacter{2085}{\textsubscript{5}}
\DeclareUnicodeCharacter{2086}{\textsubscript{6}}
\DeclareUnicodeCharacter{2087}{\textsubscript{7}}
\DeclareUnicodeCharacter{2088}{\textsubscript{8}}
\DeclareUnicodeCharacter{2089}{\textsubscript{9}}
\DeclareUnicodeCharacter{03B1}{$\alpha$}
\DeclareUnicodeCharacter{03B2}{$\beta$}
\DeclareUnicodeCharacter{03B3}{$\gamma$}
\DeclareUnicodeCharacter{03B4}{$\delta$}
\DeclareUnicodeCharacter{03BC}{$\mu$}
\usepackage{graphicx}
\usepackage{amsmath}
\usepackage{hyperref}
\usepackage{authblk}
\usepackage{booktabs}
\usepackage{float}
\usepackage[margin=2.5cm]{geometry}
\usepackage{authblk}  

\title{Oxygen K-edge X-ray Absorption Spectroscopy Database for NMC811 Layered Cathode Materials}
\author[1]{Jian He}
\author[2]{Bowen Xiu}
\author[2]{Feng Ryan Wang}
\author[1]{Frank MF de Groot}
\author[1]{Nongnuch Artrith\thanks{Corresponding author: \href{mailto:n.artrith@uu.nl}{n.artrith@uu.nl}}}
\affil[1]{Debye Institute for Nanomaterials Science, Utrecht University, Universiteitsweg 99, 3584 CG Utrecht, The Netherlands}
\affil[2]{Department of Chemical Engineering, University College London, Torrington Place, London WC1E 7JE, UK}
\date{11 August 2026}

\begin{document}

\maketitle

\begin{abstract}
X-ray absorption spectroscopy (XAS) probes the local chemical environment of the absorbing atom and is one of the most powerful characterization techniques for battery materials.
Here we present a database of simulated oxygen K-edge XAS spectra for the layered cathode material LiNi$_{0.8}$Mn$_{0.1}$Co$_{0.1}$O$_2$ (NMC811), built on the atomic structures of our recent work\cite{heDirectSimulationLiNi08Mn01Co01O22026}.
All spectra were obtained using the excited electron and core-hole (XCH) method with the R2SCAN meta-GGA functional, as implemented in the Vienna Ab initio Simulation Package (VASP).
The database covers benchmark binary oxides (TiO, Ti$_2$O$_3$, TiO$_2$, Mn$_3$O$_4$, Mn$_2$O$_3$, MnO$_2$) together with a realistic NMC811 supercell containing 60 transition metal sites at three states of charge.
Because each spectrum is resolved at the level of individual oxygen sites, the database links O K-edge spectral features to specific oxygen environments defined by their local coordination and transition metal neighbors.
All data are freely available and can serve as a reference for spectral fingerprinting, for direct comparison with experiments, and as training data for machine learning models.
\end{abstract}

\section{Background \& Summary}

The layered lithium transition metal oxides with the general formula LiMO$_2$ (M = Ni, Mn, Co) are among the most widely used cathode materials for lithium-ion batteries because of their high energy density, good rate capability, and long cycle life~\cite{goodenoughChallengesRechargeableLi2010}.
Although Ni-rich compositions such as LiNi$_{0.8}$Mn$_{0.1}$Co$_{0.1}$O$_2$ (NMC811) have attracted intense interest due to their high specific capacity ($>$200~mAh/g)~\cite{liHighnickelLayeredOxide2020}, the complex interplay between Ni, Mn, and Co occupying the same crystallographic site, combined with the mixed oxidation states of Ni, gives rise to intricate local electronic structures that are challenging to characterize experimentally.

\begin{figure}[H]
    \centering
    \includegraphics[width=0.85\textwidth]{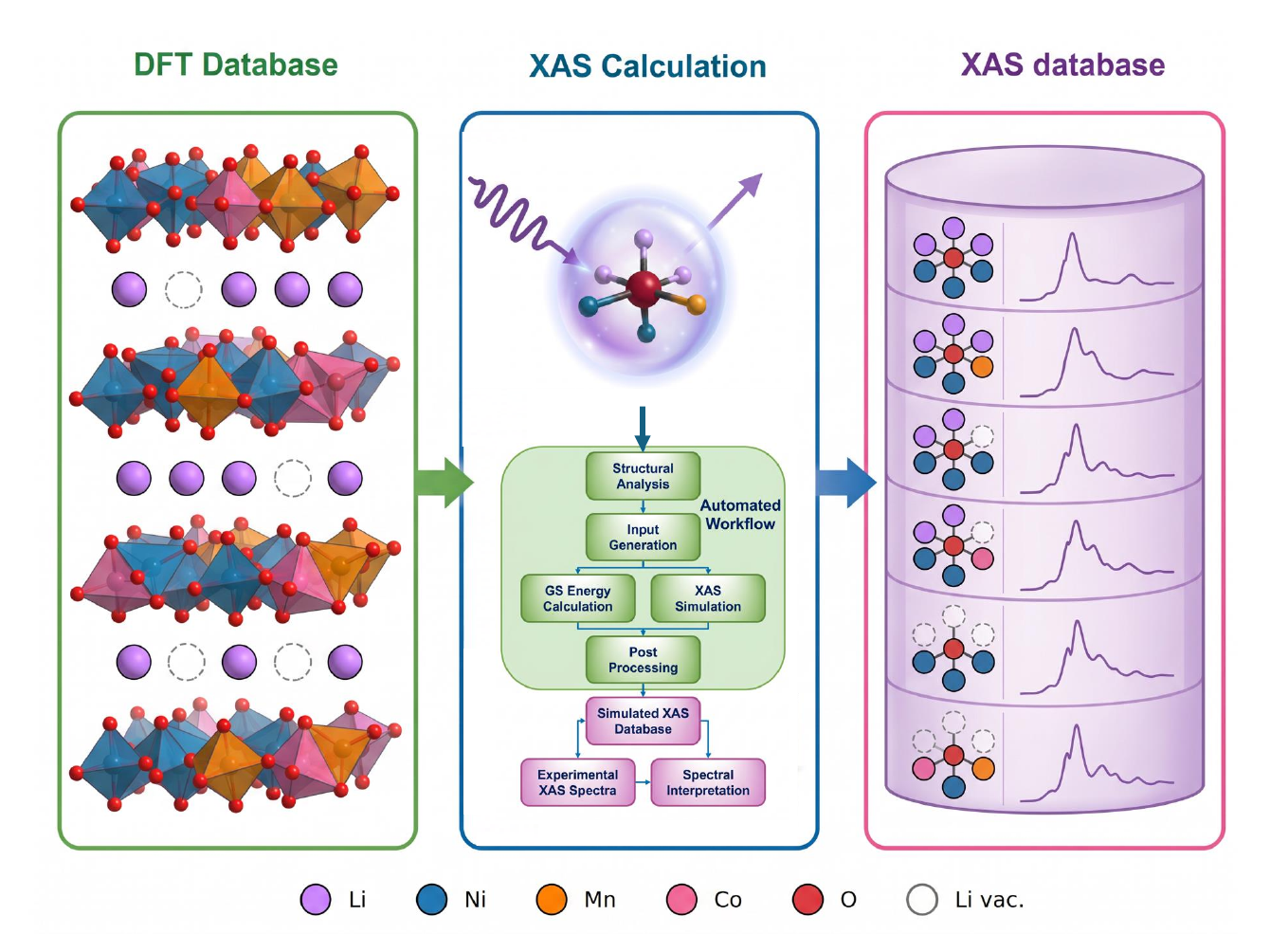}
    \caption{Flowchart with the workflow for building the oxygen K-edge XAS spectral database of NMC811 and related transition metal oxide materials. Starting from a DFT dataset of NMC811 structures, near ground-state configurations are identified. The excited electron and core-hole (XCH) calculations are then performed, in which a core electron is excited from the occupied O 1s orbital into unoccupied states and the continuum. The resulting site-resolved spectra are collected into the XAS database.}
    \label{fig:workflow}
\end{figure}

Figure~\ref{fig:workflow} illustrates the computational workflow developed for constructing the oxygen K-edge XAS spectral database.
Starting from a DFT dataset of NMC811, near ground-state configurations are identified through structural relaxation, symmetrically inequivalent oxygen sites are determined, and XCH calculations are performed for each distinct O site.
The resulting site-resolved spectra are collected into the XAS database.

A broad range of experimental probes has been applied to access the local electronic structure of NMC-type cathodes and its electrochemical consequences, from X-ray diffraction (XRD)~\cite{markerEvolutionStructureLithium2019} and nuclear magnetic resonance (NMR)~\cite{markerEvolutionStructureLithium2019} to X-ray photoelectron spectroscopy (XPS) and X-ray absorption spectroscopy (XAS)~\cite{degrootHighResolutionXrayEmission2001,degrootCoreLevelSpectroscopy2008}.
As the oxygen K-edge corresponds to transitions from the O 1s core level to unoccupied states with O 2p character, the pre-edge features directly reflect the hybridization between O 2p and transition metal 3d states~\cite{degrootOxygen1Xrayabsorption1989}.
The O K-edge XAS spectrum therefore encodes information about the oxidation state, spin state, and local coordination geometry of the transition metal ions, making it a natural choice to probe the local geometric and electronic structures in layered oxide cathodes~\cite{fratiOxygenKedgeXray2020}.

In practice, XAS spectra are most often assigned by matching their characteristic features to those of well-defined reference compounds; this strategy breaks down when no suitable reference exists or when the composition and structure of the probed material are not known a priori.
First-principles XAS simulations offer a complementary route to interpretation~\cite{taillefumierXrayAbsorptionNearedge2002,prendergastXRayAbsorptionSpectra2006}: they model explicitly how a core electron is promoted into the conduction states while a core hole is left behind.
Two families of band-structure methods within density functional theory (DFT) are commonly employed to capture the resulting final-state effect: the excited electron and core-hole (XCH) method, in which the valence electrons relax self-consistently around the core hole~\cite{taillefumierXrayAbsorptionNearedge2002,prendergastXRayAbsorptionSpectra2006,gougoussisFirstprinciplesCalculationsXray2009} (available, e.g., in XSPECTRA~\cite{giannozziQUANTUMESPRESSOModular2009} and in the Vienna Ab Initio Simulation Package, VASP~\cite{karsaiEffectsElectronphononCoupling2018,liangAccurateXRaySpectral2017}), and many-body perturbation theory based on the Bethe--Salpeter equation, in which the valence-electron screening is treated within linear response~\cite{vinsonBetheSalpeterEquationCalculations2011,gulansExcitingFullpotentialAllelectron2014}.
The latter is considerably more expensive, whereas XCH calculations reproduce experimental trends with useful accuracy at a fraction of the cost~\cite{taillefumierXrayAbsorptionNearedge2002,prendergastXRayAbsorptionSpectra2006,hamannAbsoluteApproximateCalculations2002,pueyobellafontPredictingCoreLevel2017}, which makes the XCH method well suited for building large spectral databases.
Guo et al.\ followed this route to construct a sulfur K-edge database of lithium thiophosphate solid electrolytes~\cite{guoSimulatedSulfurKedge2023}.

Computational O K-edge XAS has so far been addressed either through large, general-purpose databases based on FEFF multiple-scattering theory~\cite{mathewHighthroughputComputationalXray2018}, or, more recently, through DFT calculations of the pristine layered end-members LiCoO$_2$, LiNiO$_2$, and LiMnO$_2$~\cite{rameshAtomisticInterpretationOxygen2024}. However, no first-principles O K-edge XAS database has been reported for a realistic, multi-component disordered Ni-rich cathode such as NMC811, nor across its states of charge. Here, we address this gap by reporting site-resolved O K-edge spectra for NMC811 supercells at three states of charge, computed with the excited electron and core-hole (XCH) approach and the R2SCAN functional.
To this end, we implemented an automated XCH workflow (Fig.~\ref{fig:workflow}) that combines the open-source Pymatgen package with VASP~\cite{kresseEfficiencyAbinitioTotal1996,karsaiEffectsElectronphononCoupling2018} and employs the R2SCAN meta-GGA functional~\cite{furnessAccurateNumericallyEfficient2020}.
R2SCAN was chosen over the local density approximation (LDA) and standard GGA functionals because it describes the electronic structure and energetics of transition metal oxides more reliably~\cite{kingsburyPerformanceComparison$r^2mathrmSCAN$2022}.
The resulting database comprises nine realistic NMC811 supercell structures with 60 transition metal sites each, covering three states of charge.
Both the database and the workflow are made openly available so that other researchers can build on them directly.

\section{Methods}

\subsection{Density functional theory calculations}

All DFT calculations employed the projector-augmented-wave (PAW) method~\cite{blochlProjectorAugmentedwaveMethod1994,kresseUltrasoftPseudopotentialsProjector1999} as implemented in VASP~\cite{kresseEfficientIterativeSchemes1996}.
Numerical convergence was verified separately for the plane-wave cut-off energy, the supercell size entering the XCH calculations, the $k$-point density, and the number of unoccupied bands; in addition, different exchange-correlation functionals (Fig.~\ref{fig:functional}) and core-hole occupations (full versus zero core hole) were benchmarked.
The settings summarized below produced converged spectra in best agreement with experiment.

Both the ground-state and the core-excited calculations employed the R2SCAN meta-GGA exchange-correlation functional~\cite{furnessAccurateNumericallyEfficient2020}, which describes band gaps, magnetic moments, and the relative energetics of transition metal oxides more accurately than LDA or PBE~\cite{kingsburyPerformanceComparison$r^2mathrmSCAN$2022}, in combination with the GW-type PAW data sets from the VASP library, whose improved high-energy scattering properties benefit the description of the region above the absorption edge.
Every calculation used a plane-wave cut-off of 400~eV, a full core hole, and supercells with edge lengths of at least 10~\AA{} in each direction.
The number of bands was set to include three times as many unoccupied as occupied bands, which converges the conduction states over the energy window relevant for the O K-edge.
Occupations were smeared with a Gaussian of 0.05~eV width, and total energies were converged to better than $10^{-5}$~eV/atom.
$\Gamma$-centered $k$-point meshes were generated with Pymatgen's automatic density scheme at a uniform density of 3000~$k$-points per atom, so that all cells share a common reciprocal-space resolution (e.g., a $2\times2\times2$ mesh for the 240-atom supercells).
A constant Lorentzian broadening of 0.05~eV was applied within the XCH calculations themselves; the additional broadening needed for comparison with experiment is added in post-processing, as described below.

The XCH approach treats the final state self-consistently in the presence of a core hole in the O 1s shell~\cite{taillefumierXrayAbsorptionNearedge2002,gougoussisIntrinsicChargeTransfer2009}, and the spectrum is obtained from the imaginary part of the frequency-dependent dielectric matrix, averaged over its diagonal elements, within the PAW frozen-core approximation.
Before comparison with measured spectra, the calculated spectra are convoluted twice: with a Gaussian of 0.5~eV full width at half maximum, representing the instrumental resolution, and with a Lorentzian of energy-dependent width $0.59 + a \times (E_c - E_{\mathrm{cbm}})$~eV, representing the core-hole and quasiparticle lifetime broadening; here $a = 0.1$ is a fitting parameter, and $E_c$ and $E_{\mathrm{cbm}}$ denote the DFT conduction-band levels and the conduction-band minimum.
The absolute energy of each absorption edge is fixed by its excitation onset, computed as the total-energy difference between the core-excited final state and the ground state~\cite{hamannAbsoluteApproximateCalculations2002,englandHydrationHydrolysisCarbon2011}.

\subsection{Structure selection}

The structures for the XAS simulations comprise two categories: (\textit{i}) benchmark binary transition metal oxides and (\textit{ii}) a realistic NMC811 supercell.

The benchmark oxides were selected to span a range of transition metal oxidation states relevant to NMC811: TiO (Ti$^{2+}$), Ti$_2$O$_3$ (Ti$^{3+}$), TiO$_2$ (Ti$^{4+}$), Mn$_3$O$_4$ (Mn$^{2+}$/Mn$^{3+}$), Mn$_2$O$_3$ (Mn$^{3+}$), and MnO$_2$ (Mn$^{4+}$).
Crystal structures were obtained from the Inorganic Crystal Structure Database (ICSD) and relaxed with R2SCAN prior to the XCH calculations.

The NMC811 supercells were constructed from the layered $R\bar{3}m$ ($\alpha$-NaFeO$_2$-type) structure of LiCoO$_2$ by substituting Co with Ni and Mn on the transition metal sublattice, yielding 48 Ni, 6 Mn, and 6 Co atoms together with 120 oxygen atoms and thereby capturing the compositional disorder characteristic of the real material.
The cation configurations were drawn from the DFT dataset of our recent work~\cite{heDirectSimulationLiNi08Mn01Co01O22026}, from which the configurations lowest in energy were selected.
To sample the effect of state of charge, the supercell was considered at three lithiation levels (pristine $x=0.0$, half-delithiated $x=0.5$, and fully delithiated $x=1.0$ in Li$_{1-x}$Ni$_{0.8}$Mn$_{0.1}$Co$_{0.1}$O$_2$), with three low-energy configurations per state of charge, giving nine NMC811 structures in total.

\subsection{Automated DFT workflow for constructing XAS database}

Using the parameters established on the benchmark systems, the XCH calculations were automated in a workflow that generates the database entries (Fig.~\ref{fig:workflow}).
For every optimized structure, the symmetry-inequivalent O sites and their multiplicities are identified automatically with the symmetry tools of the Pymatgen package.
The same infrastructure then constructs the supercells and writes the VASP input files, both for the single-point R2SCAN calculation that provides the ground-state energy and for one XCH calculation per symmetry-distinct O atom of the supercell.
Post-processing of the finished calculations proceeds in two steps: the spectra of the individual O sites are first aligned via their total-energy-difference excitation onsets, and then combined in a multiplicity-weighted average that yields the spectrum of the entire structure.
The unaveraged, site-resolved spectra are retained as well: they encode the spectral signatures of the individual local atomic environments and can be readily used, for instance for machine-learning-assisted spectral interpretation.

For each structure, we provide the relaxed crystal structure, VASP input files (INCAR, POSCAR, KPOINTS) for both the ground-state and XCH calculations, and the post-processed XAS spectra.
Site-resolved spectra for each inequivalent O site are included alongside the site-averaged total spectrum.

\subsection{Sample preparation and XAS measurements}

Oxygen K-edge NEXAFS spectra were measured for a series of standard oxide samples. All standard samples were measured at Beamline B07b at Diamond Light Source using the ambient-pressure NEXAFS end station. The spectra were collected at room temperature (295~K) under pressures ranging from $10^{-7}$ to $10^{-6}$~mbar. Powder samples were spread onto carbon tape, and the sample holder was rotated to an angle of $30^\circ$ during data acquisition.
Beamline information: The B07b end station covers an energy range of 45--2200~eV with a resolving power ($E/\Delta E$) greater than 5000. The minimum beam size at the sample (FWHM) is $\leq 200~\mu$m $\times$ $\leq 200~\mu$m, and the photon flux at the sample exceeds $1 \times 10^{10}$~photons/s.

\section{Data Records}

Table~\ref{tab:database} summarizes the contents of the database, which spans the benchmark oxides and the NMC811 supercells.
Every database entry consists of one ground-state self-consistent field (SCF) calculation for the structure, plus one core-hole calculation --- performed with the O core-hole potential --- for each of its symmetry-inequivalent O sites.
For every VASP calculation, we provide the complete set of input files (INCAR, POSCAR, and KPOINTS); only the pseudopotential (POTCAR) files are omitted, as they are distributed with the VASP license, so that users can regenerate them and rerun any calculation.
Because the full VASP output would be prohibitively large, only the files needed to document and reproduce the database are retained: the three input files together with the OSZICAR file, which records the convergence behavior of each run.
Where relevant, the Fermi energy is stored as well, along with the post-processed XAS spectrum.
Each spectrum file holds four columns: the energy axis and the three Cartesian polarization components of the absorption intensity.

\begin{table}[H]
\centering
\caption{Contents of the O K-edge XAS database: number of compositions, structures, and computed O sites per category.}
\label{tab:database}
\begin{tabular}{lccc}
\toprule
\textbf{Category} & \textbf{Compositions} & \textbf{Structures} & \textbf{O Sites} \\
\midrule
Benchmark oxides & 6 & 6 & 47 \\
NMC811 & 1 & 9 & 1080 \\
\midrule
Total & 7 & 15 & 1127 \\
\bottomrule
\end{tabular}
\end{table}

The database comprises the following calculated O K-edge XAS spectra:

\begin{itemize}
    \item \textbf{Benchmark oxides:} TiO, Ti$_2$O$_3$, TiO$_2$ (rutile), Mn$_3$O$_4$, Mn$_2$O$_3$, and MnO$_2$ --- used for validation against experimental spectra.
    \item \textbf{NMC811:} A full supercell with 60 transition metal sites (48 Ni, 6 Mn, 6 Co) and 120 oxygen sites, computed at three states of charge (pristine $x=0.0$, half-delithiated $x=0.5$, and fully delithiated $x=1.0$ in Li$_{1-x}$Ni$_{0.8}$Mn$_{0.1}$Co$_{0.1}$O$_2$), with three configurations per state of charge (9 structures in total).
\end{itemize}

The XCH calculations for the NMC811 supercells were performed on the GPU partition of the Dutch national supercomputer Snellius (SURF), using the OpenACC GPU port of VASP~6.4.2 with each calculation running on a single NVIDIA H100 GPU and 16 CPU threads.
One self-consistent core-hole calculation for a 180--240-atom NMC811 supercell requires a wall time of typically 4--12 hours (median $\approx$6~hours), depending on the state of charge and the oxygen site.
With 120 core-hole calculations plus one ground-state calculation per structure, a single NMC811 structure amounts to roughly 700--1200 GPU-hours, and the complete NMC811 dataset to approximately 8,000 GPU-hours.
The benchmark oxide calculations, which involve much smaller cells, were performed on conventional CPU nodes.

Post-processing scripts that extract the key observables (energies and spectral intensities) and assemble the site-averaged spectrum of each material accompany the data.
The complete database is hosted on the Materials Cloud repository (\href{https://doi.org/10.24435/materialscloud:xx-xx}{https://doi.org/10.24435/materialscloud:xx-xx}).

\section{Technical Validation}

\subsection{Choice of the exchange-correlation functional}

\begin{figure}[H]
    \centering
    \includegraphics[width=0.52\textwidth]{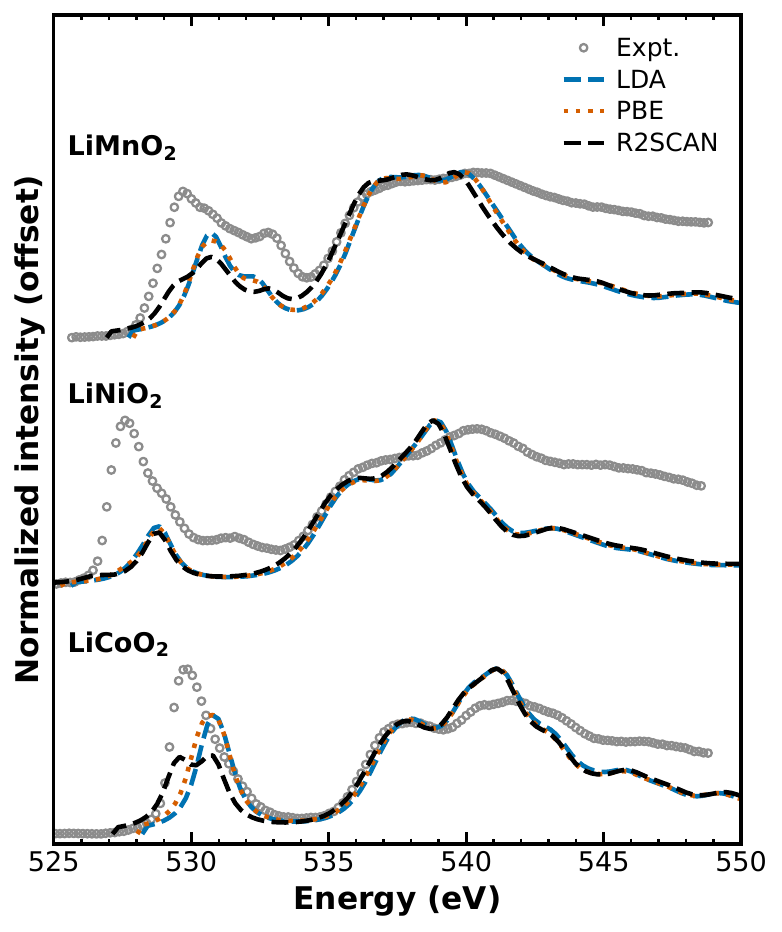}
    \caption{O K-edge XCH spectra of the layered oxide end-members computed with the LDA (blue dashed), PBE (orange dotted), and R2SCAN (black dashed) exchange-correlation functionals, using identical full-core-hole supercells, $k$-point meshes, and numbers of bands (ENCUT = 400~eV; GW-type PAW potentials of the corresponding functional family), compared with experimental spectra (grey open circles) digitized from Ramesh et al.~\cite{rameshAtomisticInterpretationOxygen2024}.
    Each calculated spectrum is aligned to the experimental pre-edge maximum of its compound; all curves are normalized to their display-window maxima, and the three compounds are vertically offset.
    R2SCAN (black) is the functional adopted for the database.}
    \label{fig:functional}
\end{figure}

To assess the sensitivity of the simulated spectra to the exchange-correlation functional, the O K-edge XCH spectra of the layered oxide end-members LiCoO$_2$, LiNiO$_2$, and LiMnO$_2$ were computed with LDA, PBE, and R2SCAN under otherwise identical conditions --- the same full-core-hole supercells, $k$-point meshes, and numbers of bands as the production settings --- and compared with the experimental spectra of Ramesh et al.~\cite{rameshAtomisticInterpretationOxygen2024} (Fig.~\ref{fig:functional}).
LDA and PBE yield nearly indistinguishable spectra for all three compounds, and R2SCAN follows them closely for LiNiO$_2$ and LiMnO$_2$.
The experimental pre-edge positions and line shapes are well reproduced, including the low pre-edge energy of LiNiO$_2$ and the split pre-edge region of LiMnO$_2$; the calculated main-edge features lie 1--2~eV closer to the pre-edge than in experiment, and the relative pre-edge intensity is underestimated, which we attribute in part to the single-particle XCH approximation and to the normalization of the digitized experimental data.
For LiCoO$_2$, R2SCAN splits the pre-edge into a doublet that is present neither in the experimental spectrum nor in the LDA and PBE calculations; this splitting is robust with respect to the magnetic initialization of the core-hole calculation.
R2SCAN was nevertheless adopted for the database because of its improved description of the ground-state electronic structure and energetics of transition metal oxides~\cite{kingsburyPerformanceComparison$r^2mathrmSCAN$2022}; the comparison in Fig.~\ref{fig:functional} quantifies the spectral uncertainty associated with this choice.

\subsection{Benchmark of the XAS simulations}

\begin{figure}[H]
    \centering
    \includegraphics[width=\textwidth]{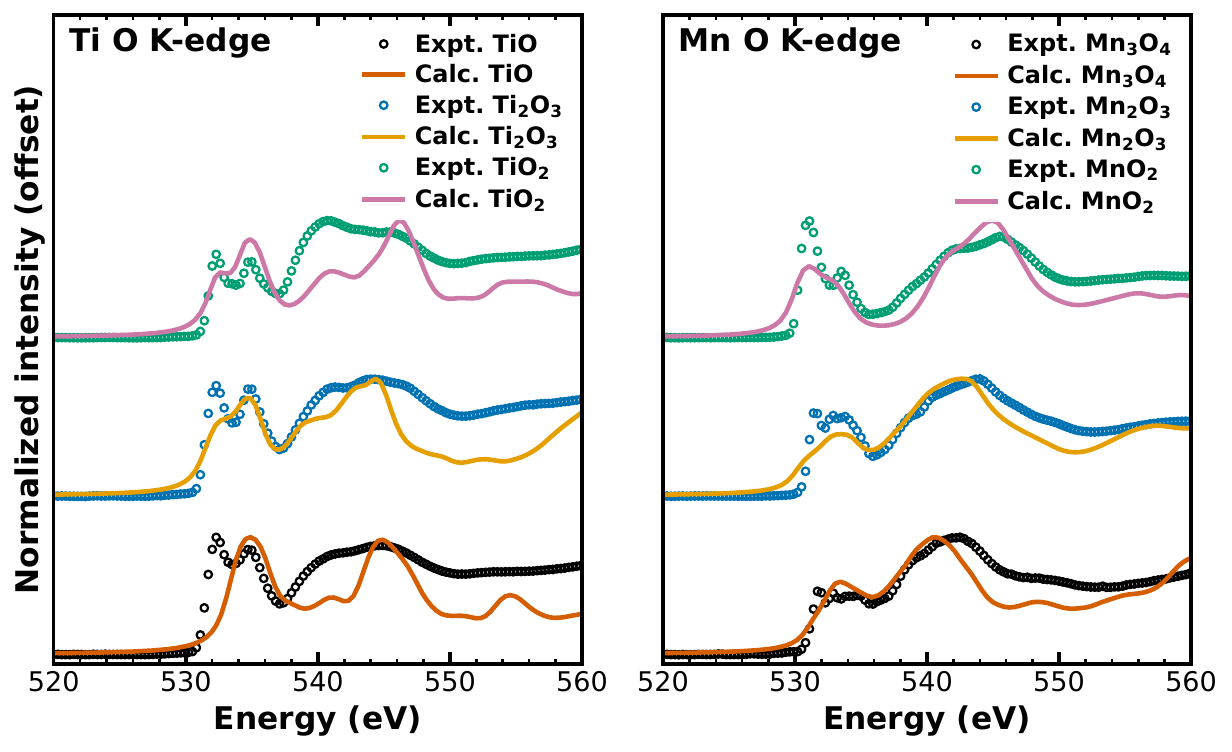}
    \caption{Benchmark results of XAS simulations of Ti oxides (left panel: TiO, Ti$_2$O$_3$, TiO$_2$) and Mn oxides (right panel: Mn$_3$O$_4$, Mn$_2$O$_3$, MnO$_2$). In each panel, the open circles indicate the experimental spectra (after subtraction of a linear pre-edge background) and the solid lines indicate the simulated spectra. The energy range covers 520--560~eV.}
    \label{fig:benchmark}
\end{figure}

As a first validation step, the XAS simulations were benchmarked on a series of binary transition metal oxides whose oxidation states bracket those encountered in NMC811: TiO (Ti$^{2+}$), Ti$_2$O$_3$ (Ti$^{3+}$), TiO$_2$ (Ti$^{4+}$), Mn$_3$O$_4$ (Mn$^{2+}$/Mn$^{3+}$), Mn$_2$O$_3$ (Mn$^{3+}$), and MnO$_2$ (Mn$^{4+}$).
Figure~\ref{fig:benchmark} compares the simulated O K-edge spectra with our experimental measurements for all six compounds; the simulations reproduce the main spectral features of every reference system.

For the Ti oxide series, the XCH simulations successfully capture the systematic evolution of the pre-edge features with increasing Ti oxidation state: the sharp pre-edge peaks corresponding to transitions into Ti 3d--O 2p hybridized states shift and change in relative intensity as the d-electron count decreases from d$^2$ (TiO) to d$^0$ (TiO$_2$), in line with the trends established for transition metal oxides by de Groot et al.~\cite{degrootOxygen1Xrayabsorption1989}.
The post-edge region, which reflects transitions to higher-lying states with mixed O 2p and Ti 4sp character, is also well reproduced.

For the Mn oxides, the calculations correctly distinguish the spectral signatures of the different Mn oxidation states present in Mn$_3$O$_4$ (mixed Mn$^{2+}$/Mn$^{3+}$), Mn$_2$O$_3$ (Mn$^{3+}$), and MnO$_2$ (Mn$^{4+}$), including the differences in pre-edge peak splitting and intensity that arise from the different d-electron configurations and crystal field environments.

Taken together, the level of agreement with experiment across the benchmark series supports the reliability of the chosen computational protocol.


\subsection{Validation of the calculated NMC811 spectra against experiment}

\begin{figure}[H]
    \centering
    \includegraphics[width=0.7\textwidth]{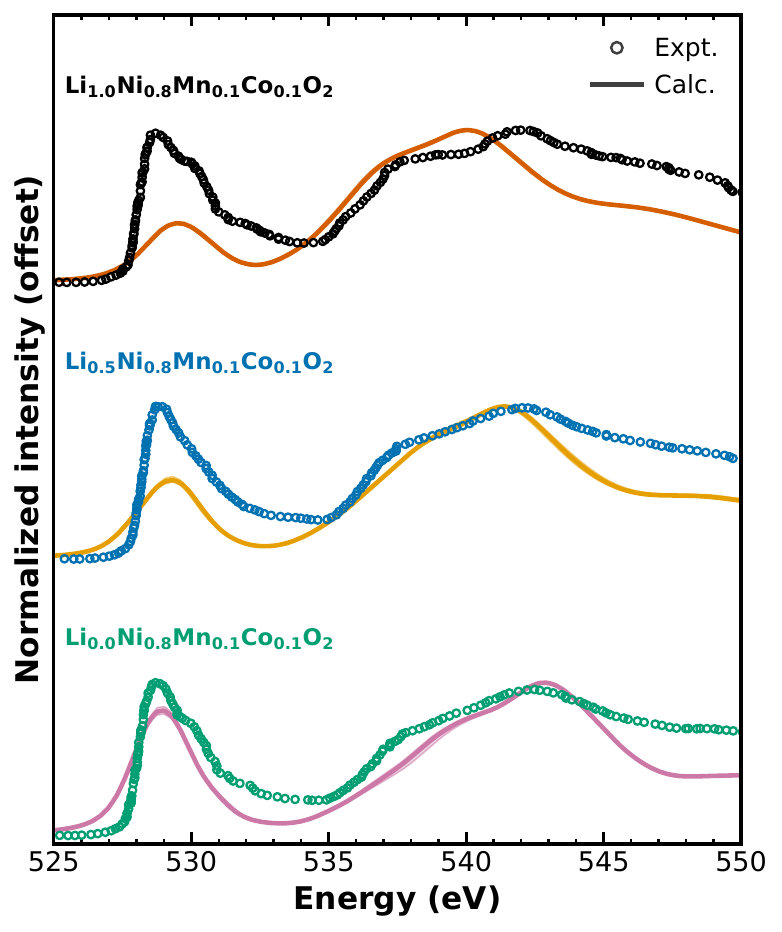}
    \caption{Calculated (solid lines) and experimental\cite{kleinerOriginReversibleIrreversible2021} (open circles) O K-edge XAS spectra of NMC811 at three states of charge, overlaid, vertically offset for clarity, and aligned on the dominant pre-edge peak. From top to bottom: pristine Li$_{1.0}$Ni$_{0.8}$Mn$_{0.1}$Co$_{0.1}$O$_2$ (calc.) paired with the discharged electrode (expt.); half-delithiated Li$_{0.5}$Ni$_{0.8}$Mn$_{0.1}$Co$_{0.1}$O$_2$ paired with 100~mAh~g$^{-1}$; and fully delithiated Li$_{0.0}$Ni$_{0.8}$Mn$_{0.1}$Co$_{0.1}$O$_2$ paired with 150~mAh~g$^{-1}$. For each state of charge the calculated spectrum is averaged over the individual configurations, and the shaded band indicates the configurational min--max spread. The energy range covers 525--550~eV.}
    \label{fig:nmc_expt}
\end{figure}

To validate the simulated XAS spectra of NMC811, we compared the calculated O K-edge spectra at three states of charge with experimental measurements at the corresponding states of charge, as shown in Fig.~\ref{fig:nmc_expt}.
The calculated spectra reproduce the key features of the experimental O K-edge and capture their evolution with delithiation.

The XCH simulations successfully reproduce the pre-edge peak near 528--530~eV arising from transitions into unoccupied O 2p--transition metal 3d hybridized states, and the broad main-edge feature centered around 540~eV corresponding to transitions into O 2p--metal 4sp hybridized states.
The pre-edge peak position is well captured by the calculation, confirming that the R2SCAN functional provides a reasonable description of the d--p hybridization in NMC811.
Both the calculated and experimental spectra show a systematic growth of the pre-edge intensity with increasing delithiation, consistent with the creation of additional unoccupied O 2p--metal 3d states as the transition metals are oxidized upon Li removal.
Some differences are observed in the relative intensities of the pre-edge and main-edge features, as well as in the depth of the valley between them ($\sim$532--536~eV), which may be attributed to the limitations of the XCH single-particle approximation, thermal disorder effects not captured in the static supercell calculation, and possible surface contributions in the experimental spectrum.
The overall agreement indicates that the DFT simulations capture the essential physics of the O K-edge in NMC811 and supports the reliability of the database.

\subsection{Oxidation state analysis from magnetic moments}

\begin{figure}[H]
    \centering
   \includegraphics[width=1.07\textwidth]{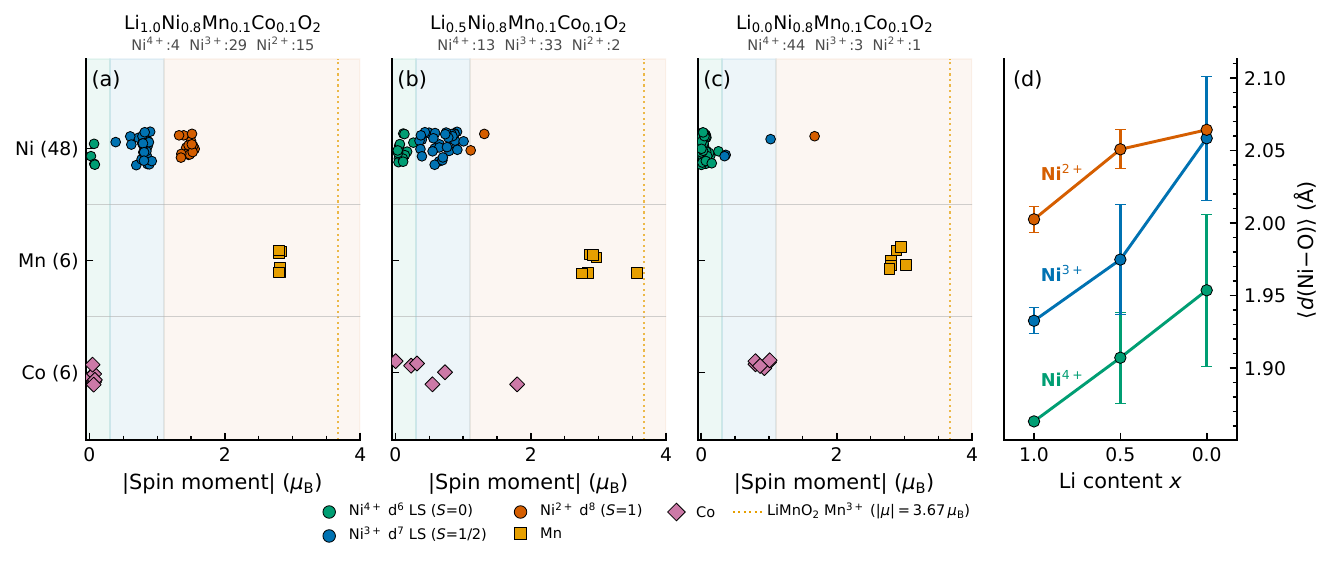}
    \caption{\textbf{Distribution of the absolute spin moments of the transition metal ions (Co, Mn, Ni) in the NMC811 supercell at three states of charge}: \textbf{(a)} pristine Li$_{1.0}$Ni$_{0.8}$Mn$_{0.1}$Co$_{0.1}$O$_2$, \textbf{(b)} half-delithiated Li$_{0.5}$Ni$_{0.8}$Mn$_{0.1}$Co$_{0.1}$O$_2$, and \textbf{(c)} fully delithiated Li$_{0.0}$Ni$_{0.8}$Mn$_{0.1}$Co$_{0.1}$O$_2$.
    One representative configuration is shown per state of charge.
    The counts of Ni in each oxidation state (Ni$^{4+}$/Ni$^{3+}$/Ni$^{2+}$) are annotated in each panel.
    The computed moments correspond to low-spin configurations throughout: Ni$^{4+}$ (d$^6$, $S=0$), Ni$^{3+}$ (d$^7$, $S=1/2$), and Ni$^{2+}$ (d$^8$, $S=1$); no Ni site approaches the $\sim$2.6~$\mu_\mathrm{B}$ expected for high-spin Ni$^{3+}$.
    \textbf{(d)} Average Ni--O bond length for each Ni oxidation state (mean $\pm$ standard deviation over the Ni sites of each class), providing an independent geometric confirmation of the moment-based classification.}
    \label{fig:magmom}
\end{figure}

The oxidation states of the transition metals in the relaxed NMC811 supercell were determined from the calculated magnetic moments, as shown in Fig.~\ref{fig:magmom}.
For each state of charge, one representative configuration out of the three is analyzed.
The analysis reveals several important features of the electronic structure of NMC811 and its evolution with state of charge.

In the pristine supercell, Co remains in the low-spin Co$^{3+}$ (d$^6$) configuration with magnetic moments near zero, indicating a completely filled $t_{2g}$ shell, and Mn adopts the Mn$^{4+}$ (d$^3$) oxidation state with magnetic moments of $\sim$2.9~$\mu_\mathrm{B}$, significantly lower than the $\sim$3.7~$\mu_\mathrm{B}$ expected for Mn$^{3+}$.
Ni shows a broad distribution of magnetic moments spanning three oxidation states: Ni$^{2+}$ (d$^8$, $\sim$1.5~$\mu_\mathrm{B}$), Ni$^{3+}$ (d$^7$, $\sim$0.9~$\mu_\mathrm{B}$), and Ni$^{4+}$ (d$^6$, $\sim$0.3~$\mu_\mathrm{B}$).
The moment magnitudes identify all Ni states as low-spin configurations: in particular, Ni$^{3+}$ is exclusively low-spin ($t_{2g}^6e_g^1$, $S=1/2$) --- across all three states of charge the largest computed Ni moment is 1.7~$\mu_\mathrm{B}$, far below the $\sim$2.6~$\mu_\mathrm{B}$ expected for high-spin Ni$^{3+}$ ($S=3/2$).
Upon delithiation, the Ni population shifts progressively toward higher oxidation states, with Ni$^{4+}$ becoming dominant in the fully delithiated structure, while Co and Mn remain essentially unchanged --- consistent with Ni acting as the primary redox-active center in NMC811.
The oxidation-state assignment is independently corroborated by the local geometry (Fig.~\ref{fig:magmom}d): within each structure the average Ni--O bond length decreases stepwise with increasing Ni oxidation state (in the pristine supercell, 2.00~\AA{} for Ni$^{2+}$, 1.93~\AA{} for Ni$^{3+}$, and 1.86~\AA{} for Ni$^{4+}$), so the mean Ni--O distance tracks the moment-based charge classification.

The charge disproportionation of Ni and its evolution with state of charge have important implications for the O K-edge spectral shape, as different Ni oxidation states create different unoccupied d--p hybridized states accessible to the O 1s core electron.

\subsection{Diversity of local oxygen environments}

\begin{figure}[H]
    \centering
    \includegraphics[width=0.92\textwidth]{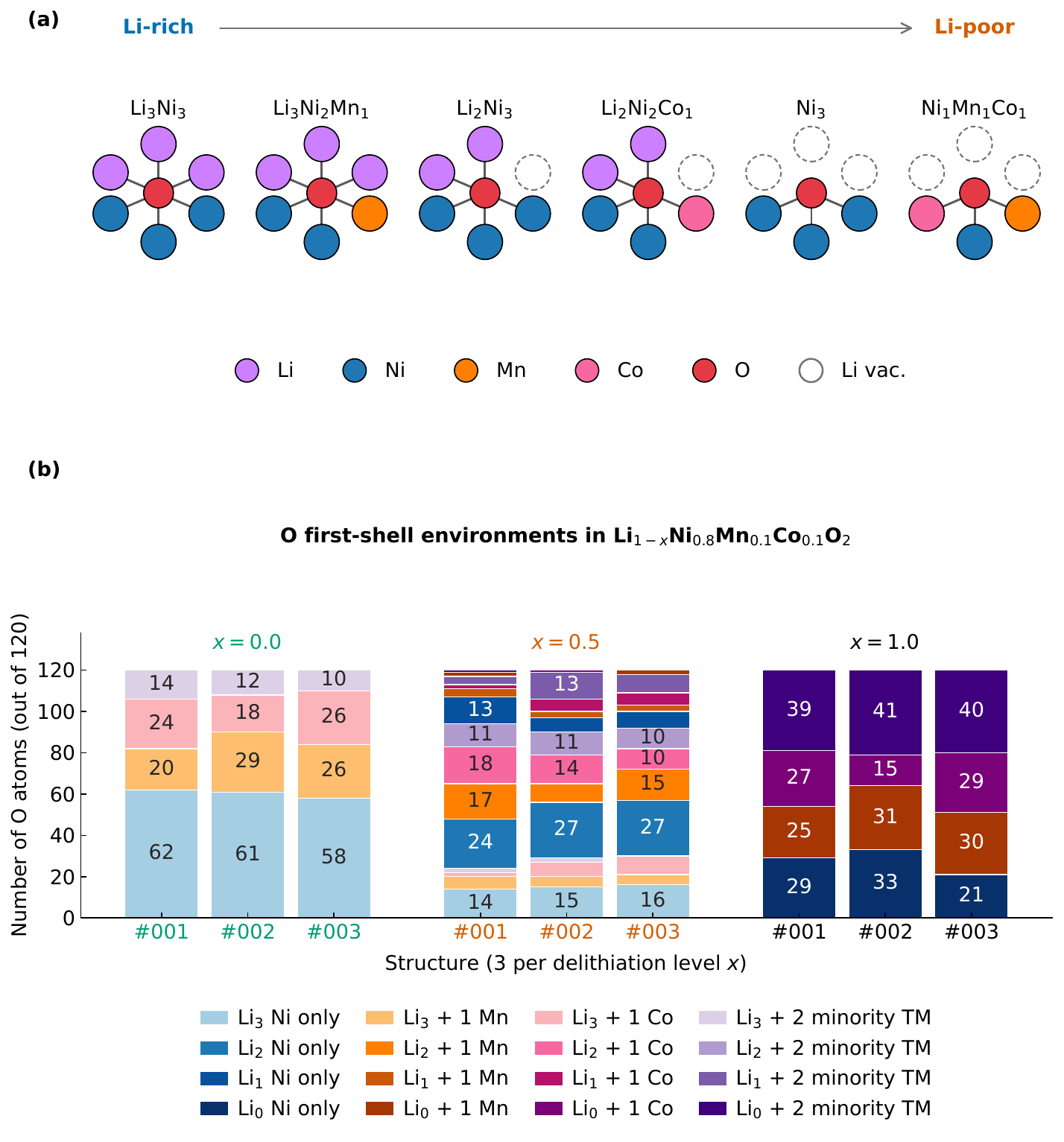}
    \caption{Local oxygen environments in Li$_{1-x}$Ni$_{0.8}$Mn$_{0.1}$Co$_{0.1}$O$_2$.
    \textbf{(a)} Idealized representative oxygen first-shell coordination motifs, ordered from Li-rich to Li-poor: each O atom (red) is coordinated by three sites of the Li layer (Li or vacancy, dashed circles) and three sites of the transition metal layer (Ni, Mn, Co).
 \textbf{(b)} Distribution of first-shell oxygen environments across the nine NMC811 structures (three configurations at each of $x=0.0$, $0.5$, and $1.0$).
    Each oxygen atom (120 per structure) is classified by the composition of its six nearest cation neighbours (Li, Ni, Mn, Co, or vacancy), and the stacked bars give the number of oxygen atoms in each environment class per structure.}
    \label{fig:oenv}
\end{figure}

A key feature of the database is that it resolves the O K-edge spectrum down to individual, symmetry-distinct oxygen sites, each of which samples a different local coordination environment.
To characterise this diversity, every oxygen atom in the nine NMC811 structures was classified by its first cation coordination shell (the six nearest cations, accounting for Li vacancies created upon delithiation), as shown in Fig.~\ref{fig:oenv}.

The classification reveals a rich spectrum of distinct oxygen environments that becomes increasingly diverse upon delithiation, as Li vacancies and the mixed Ni/Mn/Co occupation of the transition metal layer generate many inequivalent local coordinations.
This local heterogeneity is precisely what the site-resolved spectra in the database are designed to capture: each distinct environment contributes a characteristic O K-edge signature, and their superposition gives rise to the overall spectral shape of the disordered material.
The environment-resolved data therefore enable users to correlate individual spectral features with specific local oxygen coordinations, supporting applications such as spectral fingerprinting and the training of machine-learning models that map local structure to spectral response.

\section{Usage Notes}

See Code availability.

\section{Code Availability}

The automated workflow, together with the post-processing scripts used to extract spectra and energies from the VASP output, is open source and available on GitHub. (\href{https://github.com/atomisticnet/xas-tools}{https://github.com/atomisticnet/xas-tools}, release v J.H.)
The database itself is hosted on the Materials Cloud repository (\href{https://doi.org/10.24435/materialscloud:xx-xx}{https://doi.org/10.24435/materialscloud:xx-xx}).

\section{Acknowledgements}

This work was supported by a start-up grant (Dutch Sector Plan) from Utrecht University awarded to N.A. The DFT calculations were performed using the ænetone HPC infrastructure, supported by a Dutch Sector Plan grant. J.H. and N.A. thank the SURF Cooperative for providing access to the Dutch national e-infrastructure under grants no. EINF-5267 and no. EINF-18375, on which the XAS DFT core-hole calculations were performed.
We thank Diamond Light Source for access to the ambient-pressure NEXAFS end station at beamline B07b, where the experimental reference spectra were measured.

\section{Author contributions}

J.H.\ performed all XAS calculations, including the benchmarking and parameter optimization, curated the database, analysed the data, and wrote the manuscript.
B.X.\ carried out the experimental NEXAFS measurements of the benchmark oxides.
F.R.W.\ supervised the experimental work and contributed to the discussion of the results.
F.M.F.d.G.\ contributed to the interpretation of the XAS spectra and to the discussion of the results.
N.A.\ conceived and supervised the project, contributed to the discussion of the results, and revised the manuscript.
All authors reviewed and approved the final manuscript.

\section{Competing interests}

The authors declare no competing interests.

\bibliographystyle{naturemag}
\bibliography{references}
\end{document}